\documentclass[conference]{IEEEtran}
\IEEEoverridecommandlockouts
\usepackage{cite}
\usepackage{amsmath,amssymb,amsfonts}
\usepackage{graphicx}
\usepackage{textcomp}
\usepackage{xcolor}
\usepackage{booktabs}

\usepackage{placeins}
\usepackage{multirow}

\def\BibTeX{{\rm B\kern-.05em{\sc i\kern-.025em b}\kern-.08em
    T\kern-.1667em\lower.7ex\hbox{E}\kern-.125emX}}

\usepackage[hidelinks]{hyperref}
\usepackage{url}
\begin{document}
% Encoder Over Decoder?: Probing Speech LLMs and SSL Models for Cross-lingual Parkinson's Disease Detection
\title{Exploiting Speech LLM Representations for Multilingual and Cross-Lingual Parkinson's Disease Detection
\\
}

\author{
\IEEEauthorblockN{Sarthak Giri, Zi Haur Pang, Tatsuya Kawahara}
\vspace{1mm}
\IEEEauthorblockA{\textit{Graduate School of Informatics, Kyoto University, Japan}\\ [1mm]
\{giri, pang\}@sap.ist.i.kyoto-u.ac.jp,  kawahara@i.kyoto-u.ac.jp
}
}

\maketitle

\begin{abstract}
Speech Large Language Models (Speech LLMs) have shown strong performance across diverse tasks, yet their utility for pathological speech analysis remains underexplored. In this work, we investigate the effectiveness of internal representations from encoder and decoder components of Speech LLMs for Parkinson's Disease (PD) detection across multilingual and cross-lingual settings. Our findings reveal that encoder representations  consistently outperform their decoder counterparts in most models and settings and that pathological cues may be progressively attenuated as audio representations are projected into the language model space. We further show that generative outputs are less reliable for clinical tasks compared to internal representations. To leverage information spread across multiple layers, we propose a Squeeze-and-Excitation (SE)-based dynamic layer aggregation framework, which surpasses best-layer selection in multiple experiments, suggesting that PD-relevant acoustic cues are distributed across transformer layers rather than concentrated in one.

%  s largely unexplored. In this work, we first analyze Speech LLM generative prediction for Parkinson’s disease detection using both raw-audio and biomarker-guided prompts. Motivated by the limited reliability of generative prediction, we then systematically probe audio encoder and LLM decoder embeddings across Speech LLMs to understand where in the Speech LLM pathological cues are highest for cross-lingual and multilingual analysis.

% Our results reveal a consistent hierarchy of discriminative power, with audio encoder representations substantially outperforming decoder embeddings, which in turn outperform generative outputs. Layer-wise analyses suggest that pathology-relevant acoustic information is progressively attenuated as speech representations are projected into the language modeling space. Motivated by these findings, we propose a robust sample-adaptive Squeeze-and-Excitation (SE) based layer aggregation framework that dynamically combines encoder representations across layers, eliminating exhaustive best layer selection while achieving performance comparable to the single strongest individual layer. These findings provide new insights into representation learning in Speech LLMs and offer practical guidance for its application in clinical speech analysis.

\end{abstract}


\section{Introduction}
Parkinson's disease (PD) affects speech production through impairments in phonation, articulation, prosody and speech timing, making speech a promising non-invasive biomarker for disease detection and monitoring \cite{bloem2021parkinson,ackermann1991articulatory,ramig2008speech,rusz2011quantitative, skodda2011aspects}. Traditional approaches of speech analysis were based on machine learning on handcrafted biomarker features\cite{orozco2015voiced, hossain2024machine}, but speech foundation models such as HuBERT\cite{hsu2021hubert}, Whisper\cite{radford2023robust}, WavLM\cite{chen2022wavlm}, and XLS-R\cite{babu2021xls}
pretrained on large-scale audio have emerged as effective feature extractors for pathological speech analysis, with embeddings shown to outperform or complement handcrafted biomarkers for PD detection \cite{dao2025detection,gimeno2025unveiling,klempir2025ranking,sedigh2025speech, favaro2023interpretable,la2024exploiting,brueckner2025detecting}. Adapting the embeddings of these models for cross-lingual generalization has also been explored for PD detection \cite{hernandez2026adapting}. Many recent studies have also explored fine-tuning these models to adapt them for PD detection\cite{xiong2025mitigating, purohit2025automatic}.

Recently, Speech LLMs have extended foundation models with audio-language reasoning capabilities, and existing studies for PD have explored these models primarily through zero-shot prompting, using raw audio or handcrafted biomarker features as additional context\cite{kabir2026zero, ortiz2026multimodal}. However, these studies treat Speech LLMs as black-box predictors, relying solely on their generated outputs without examining whether the internal representations carry richer pathological information.

Recent analyses of Speech LLMs suggest that paralinguistic information may become less accessible as audio representations are transformed into language representations \cite{chen2026audio, pang2026erm, pang2026paralinguistic}. Since pathological speech is characterized by subtle paralinguistic abnormalities, it remains unclear whether these clinically relevant cues are preserved within Speech LLM representations.

% A key unanswered question in clinical speech analysis is whether Speech LLMs merely provide a convenient generative output, or whether their internal representations encode discriminative pathology-related information that can be exploited more reliably than generated labels. 

% While the audio encoders of these multimodal architectures are optimized for acoustic understanding, their representations are projected into the latent space of the LLM backbone through projector layers, which may suppress subtle motor-speech cues. To date, no systematic study has compared audio encoder embeddings, decoder embeddings, and generative outputs of multimodal LLM for multilingual and cross-lingual PD detection. Such a study is important for understanding where pathology-related cues are preserved within the Speech LLM pipeline and where they are weakened.

We address this problem by systematically probing Speech LLM representations for multilingual and cross-lingual PD detection. Specifically, we investigate whether pathology-related information is preserved across different stages of the Speech LLM pipeline by comparing audio encoder representations, LLM decoder representations, and generated outputs. In addition, we evaluate generative predictions using both raw-audio prompts and prompts enriched with handcrafted biomarker features alongside the audio, to assess whether explicit biomarker information provides additional context for PD detection.

Our analysis reveals that audio encoder representations consistently outperform LLM decoder representations. Additionally, within both components,  classification performance peaks at intermediate layers, consistent with prior probing studies on dysarthric speech  \cite{yue2026probing}. However, selecting a single best layer requires exhaustive search and may not generalize consistently across models or languages. To address this issue, we propose a sample-adaptive Squeeze-and-Excitation (SE)-based layer aggregation method, inspired by SENet \cite{hu2018squeeze}\cite{guragain2024speech}, which dynamically weights contributions from different layers without manual layer search. This method outperforms both the best-layer representation and the representation obtained by mean-pooling embeddings across all Transformer layers in multiple settings, suggesting PD-relevant cues are distributed unevenly across transformer layers, such that neither a single layer nor equal weighting of all layers is consistently optimal.

\begin{figure*}[!t]
\centering
% \hspace*{-1cm}
\includegraphics[width=0.99\textwidth]{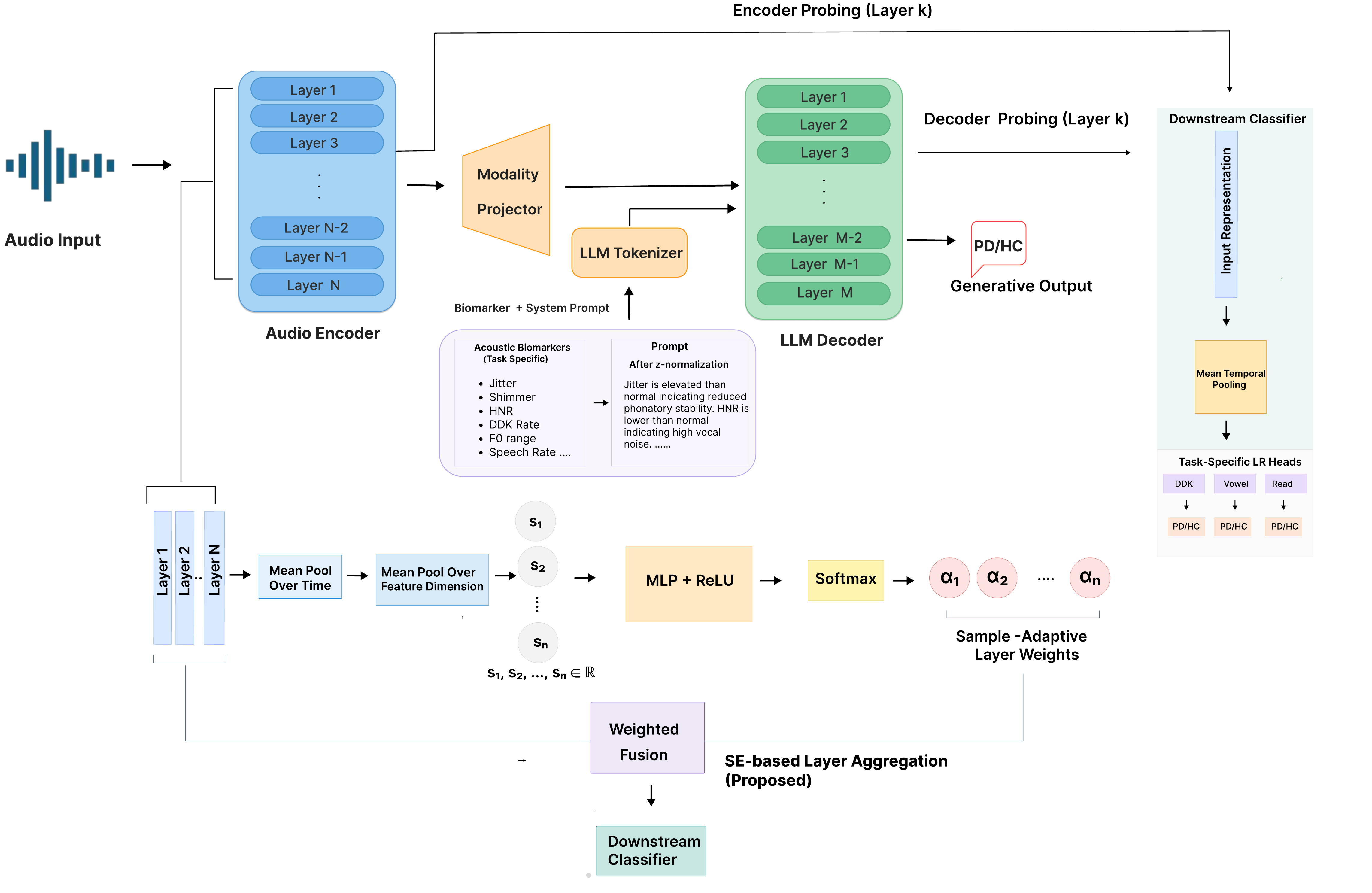}
\caption{Overview of Speech LLM pipeline showing the encoder/decoder representation probing, zero-shot generative output, and the proposed SE-based layer aggregation.}
\label{fig:arch}
\end{figure*}

The main contributions of this work are:
\begin{itemize}
    \item To our knowledge, the first systematic comparison of audio encoder representations, LLM decoder representations, and generative-output of Speech LLMs for multilingual and cross-lingual PD detection.
            \item An  empirical hierarchy, $\mathrm{encoder} > \mathrm{decoder} > \mathrm{generative\ output}$, showing that pathological cues are progressively attenuated through the Speech LLM.
    \item 
A sample-adaptive SE-based layer aggregation method that outperforms single best-layer performance in most settings, reducing the need for exhaustive layer search.
\end{itemize}

The remainder of this paper is organized as follows. Section~II describes the layer-wise representation probing approach, the proposed SE-based layer aggregation method, and the zero-shot generative evaluation. Section~III presents the experimental setup. Section~IV reports and discusses the results, and Section~V concludes the paper.

\section{Methods}

\subsection{Overview}

% \begin{figure*}[t]
%   \centering
%   \includegraphics[width=\columnwidth]{unused/Frame 15 (2).pdf}
%   \caption{Speech LLM pipeline showing the three probing pathways: encoder probing, decoder probing, and zero-shot generative output. }
%   \label{fig:arch}
% \end{figure*}

We investigate PD detection using Speech LLMs via three complementary pathways: audio encoder representations, LLM decoder representations, and zero-shot generative outputs, including biomarker-enriched prompting. Since pathological cues may be distributed across layers rather than concentrated in one, we further propose a sample-adaptive SE-based layer aggregation method that dynamically combines representations across layers. All probing experiments operate on frozen pretrained models with only lightweight downstream classifiers trained.

\subsection{Layer-Wise Representation Probing}
We perform layer-wise probing of frozen hidden representations from both encoder and decoder components of Speech LLMs. Decoder representations were extracted from a single forward pass conditioned on the audio and prompt tokens prior to autoregressive generation.
Let $\{\mathbf{H}_1, \mathbf{H}_2, \dots, \mathbf{H}_N\}$ be the outputs from each layer, where $\mathbf{H}_i \in \mathbb{R}^{T \times F}$ denotes the hidden representation at layer $i$, $T$ is the number of temporal frames for encoder representations or sequence positions for decoder representations, and $F$ is the feature dimension.

 The hidden states from layer \(i\) are mean-pooled over the corresponding temporal or sequence dimension before feeding to the classifiers.
\begin{equation}
\mathbf{z}_i = \frac{1}{T}\sum_{t=1}^{T} \mathbf{H}_i[t],
\end{equation}

Encoder probing targets acoustic representations before language-model projection; decoder probing measures how much PD-related information persists after projection (see Fig.~\ref{fig:arch}).

\subsection{SE-Based Layer Aggregation}
PD detection performance can vary across Transformer layers. Training a separate classifier per layer and
selecting the best is computationally expensive and may not
generalize across models or languages. Furthermore, the PD-related cues may not be concentrated in one optimal layer, but distributed across layers.

To address this, we propose a trainable aggregation framework over frozen encoder hidden states 
$\mathbf{H} \in \mathbb{R}^{N \times T \times F}$, 
where $N$, $T$, $F$ denote number of layers, 
time frames, and feature dimension respectively.

Each layer is first summarized by average pooling over the time and feature dimensions:

\begin{equation}
s_i = \frac{1}{TF}\sum_{t=1}^{T}\sum_{f=1}^{F} H_{i,t,f},
\quad
\mathbf{s} = [s_1, s_2, \ldots, s_N] \in \mathbb{R}^{N}.
\end{equation}

These summaries are passed through a lightweight MLP and softmax to obtain normalized layer weights (Fig.~\ref{fig:arch}):

\begin{equation}
    \mathbf{a} = \text{softmax}\!\left(
    \mathbf{W}_2\,\text{ReLU}(\mathbf{W}_1\mathbf{s})
    \right), \quad \mathbf{a} \in \mathbb{R}^{N}
\end{equation}

The fused representation is computed as:

% \begin{equation}
%     \tilde{\mathbf{H}}_{b} = \sum_{i=1}^{N} 
%     a_{b,i}\,\mathbf{H}_{b,i}
% \end{equation}

\begin{equation}
\tilde{\mathbf{H}}
=
\sum_{i=1}^{N}
a_i \mathbf{H}_i
\end{equation}
The resulting weight vector $\mathbf{a}$ is computed independently for each input sample, allowing the model to dynamically emphasize different layers depending on the acoustic characteristics of the utterance. Unlike best-layer selection, which assumes a single globally optimal layer, this method learns sample-adaptive layer importance weights.

\subsection{Zero-Shot Generative Evaluation}
We also evaluate zero-shot generative classification under two prompt conditions. The raw-audio prompt instructs the Speech LLM to base its decision solely on acoustic evidence including voice quality, articulation, and prosody. The biomarker-guided prompt additionally provides descriptions of handcrafted acoustic biomarkers as text alongside the audio.

Biomarker selection was performed using Spearman correlation between each feature and the PD/HC label computed independently for Spanish, German, and Czech. The feature set was collected from existing studies on acoustic biomarkers for PD \cite{bocklet2011detection, skodda2008speech, little2008suitability}. A feature was retained if it achieved statistical significance $(p < 0.05)$ in at least one language and showed consistent directional agreement (same correlation sign) in at least two of three languages. This yielded 17 features across three speech tasks: seven vowel-task features capturing phonatory stability (jitter, shimmer, Harmonics-to-Noise Ratio (HNR), F0 variability, Cepstral Peak Prominence Smoothed (CPPS), maximum phonation time, silence ratio), five diadochokinetic (DDK) features capturing articulatory timing and rhythm (DDK rate, inter-burst variability, DDK regularity, silence ratio, burst amplitude variability), and five read-speech features capturing prosody and fluency (speech rate, pause ratio, mean pause duration, F0 range, local pitch variability).

% \begin{figure*}[t]
%   \centering
%   \includegraphics[width=\textwidth]{SE.drawio.png}
%   \caption{Proposed SE-based layer aggregation. }
%   \label{fig:se}
% \end{figure*}

\section{EXPERIMENTAL SETUP }
\subsection{Dataset}
Experiments are conducted on three PD speech datasets in Czech\cite{riosurrego2024pdetcrosslanguage}, German\cite{bocklet2011detection}, and Spanish (the PC-GITA corpus from Colombia\cite{orozco2014new}). These datasets were provided by the organizing committee of OneVoice-MSD special session. Table \ref{tab:dataset_task_distribution} summarizes the utterance-level distribution across languages, tasks, and classes. All train, validation, and test partitions in the experiments are speaker-disjoint.

\subsection{Evaluation Protocol}
We evaluate both multilingual and leave-one-language-out (LOLO) cross-lingual settings. In the LOLO configuration, two languages are used for model training and validation, while the remaining language is held out entirely for testing. The three cross-lingual evaluation settings are German+Czech→Spanish, Spanish+Czech→German, and Spanish+German→Czech. In the multilingual training, all languages are pooled while maintaining speaker-disjoint partitions across train, validation, and test scenarios. All reported metrics are computed at the utterance level.

For zero-shot Speech LLM experiments we use the same LOLO and multilingual test data used in the representation-probing. This ensures that generative outputs, decoder probes, encoder probes, and layer aggregation results are compared on the same test sets. Representation-probing and layer-aggregation results are reported using receiver operating characteristic curve (AUC) and unweighted average recall (UAR), whereas zero-shot generative results are evaluated using UAR only from discrete textual PD/HC predictions.

Statistical significance was assessed using one-sided paired speaker-level
bootstrap tests with 10,000 resamples and Benjamini--Hochberg correction.

\begin{table}[htbp]
\centering
\caption{Utterance-level sample distribution by language, task, and class. Counts are reported as PD/HC.}
\label{tab:dataset_task_distribution}
\small
\begin{tabular}{lccc}
\toprule
\textbf{Language} &
\textbf{Sustained Vowel} &
\textbf{Read} &
\textbf{DDK} \\
\midrule
Spanish & 450/447 & 90/90 & 90/90 \\
German  & 176/175 & 88/88 & 150/88 \\
Czech   & 100/100 & 100/100 & 100/100 \\
\bottomrule
\end{tabular}

\vspace{1mm}
\end{table}

% \subsection{Statistical Significance Testing}
% Statistical significance testing was performed to determine
% whether the observed performance differences between the
% compared approaches were consistent across experimental
% settings rather than arising from random variation. One-sided
% paired Wilcoxon signed-rank tests were employed for all
% comparisons. For the encoder--decoder comparison
% (Table~\ref{tab:encoder_decoder_comparison}), the test was
% performed on 16 encoder--decoder pairs (4 Speech LLMs
% $\times$ 4 evaluation settings). For the layer aggregation
% analysis (Table~\ref{tab:se_val_selected_results}), separate
% tests were performed between SE aggregation and mean
% pooling, and between SE aggregation and validation-selected
% best-layer, each using 28 paired observations (7 models
% $\times$ 4 evaluation settings). Statistical significance was
% evaluated independently for the AUC and UAR metrics.

\subsection{Models} 
We evaluate four Speech LLMs: Qwen2-Audio \cite{chu2024qwen2}, Phi-4-Multimodal \cite{abouelenin2025phi}, Ultravox \footnote{\url{https://huggingface.co/fixie-ai/ultravox-v0_5-llama-3_2-1b}},
and Qwen2.5-Omni\cite{qwen2.5omni}. For comparison with strong speech foundation model baselines, we additionally evaluate Whisper-large, XLS-R 2B, and HuBERT-large. These models provide a reference point for determining whether Speech LLM's representations offer advantages beyond conventional speech foundation models and whether the SE-based aggregation strategy generalizes across speech foundation models.

Logistic Regression was used as the classifier and a small neural head was used for training the layer aggregation module.

\subsection{Training Strategy} 

To account for the different acoustic characteristics of DDK, read speech, and sustained vowels, we used task-specific classification heads in each experimental setting, including the three LOLO settings and the multilingual setting. For each layer, one logistic-regression classifier was trained per task using samples from the corresponding task in the training data. During evaluation, each test sample was passed to the classifier associated with its task, and the resulting predictions from all tasks were combined to compute the overall AUC and UAR for that layer.

For the SE-based aggregation experiments, a small neural aggregation module with a two-layer MLP was trained to estimate layer weights over the frozen encoder hidden states as shown in Fig.~\ref{fig:arch}. 
 These weights were applied to the layer embeddings to compute a weighted aggregated embedding. 

\subsection{Best Layer Selection}

For the layer-wise representation-probing of Speech LLMs and speech foundation models, the best layer for each model and experimental setting is defined as the layer achieving the highest test-set AUC.

For each model and experimental setting in the SE aggregation comparison, the best layer is selected using validation AUC from the training languages and then evaluated on the held-out test set. This ensures a fair comparison with the SE aggregation method, which is also trained and validated exclusively on training data.

\section{RESULTS}

\subsection{Generative Output of Speech LLMs}
Table~\ref{tab:generative_results} summarizes zero-shot generative output performance. Across all models, settings, and prompting strategies, UAR values remained near chance ($\approx$ 0.50), indicating that Speech LLMs could not reliably distinguish PD from healthy controls through direct generation. This may be because Speech LLMs were trained for general audio-language understanding rather than detecting subtle pathological speech cues.

Furthermore, biomarker-guided prompting yielded no consistent improvement over raw-audio prompting across models or languages. This could be due to several factors. Subtle acoustic cues relevant to PD may be partly lost, or become less accessible, when acoustic representations are projected into the language-model space, limiting their integration with textual biomarker descriptions. In addition, Speech LLMs are not specifically trained to jointly interpret clinical biomarkers and speech for PD detection. The selected biomarkers may also provide limited or redundant information beyond the acoustic evidence already represented by the model. These results are consistent with a prior zero-shot Speech LLM evaluation for PD detection~\cite{kabir2026zero}, suggesting that generative pathways are insufficient for reliable clinical PD detection and motivating a deeper investigation of internal representations.
\begin{table}[htbp]
\centering
\caption{UAR of zero-shot generative classification using audio-only and biomarker-guided prompts alongside audio.}
\label{tab:generative_results}
\footnotesize
\setlength{\tabcolsep}{3pt}
\begin{tabular}{llcc}
\toprule
\textbf{Model} & \textbf{Setting} &
\begin{tabular}[c]{@{}c@{}}\textbf{Audio}\\\textbf{Only}\end{tabular} &
\begin{tabular}[c]{@{}c@{}}\textbf{Biomarker}\\\textbf{Guided}\end{tabular} \\
\midrule
\multirow{4}{*}{Qwen2-Audio}
& Multi & 0.550 & 0.513 \\
& ES    & 0.558 & 0.551 \\
& CS    & 0.497 & 0.498 \\
& DE    & 0.485 & 0.473 \\
\midrule
\multirow{4}{*}{Phi-4}
& Multi & 0.540 & 0.523 \\
& ES    & 0.512 & 0.581 \\
& CS   & 0.500 & 0.518 \\
& DE    & 0.500 & 0.497 \\
\midrule
\multirow{4}{*}{Ultravox}
& Multi & 0.489 & 0.498 \\
& ES    & 0.526 & 0.507 \\
& CS    & 0.510 & 0.500 \\
& DE    & 0.529 & 0.511 \\
\midrule
\multirow{4}{*}{Qwen2.5-Omni}
& Multi & 0.500 & 0.500 \\
& ES    & 0.520 & 0.500 \\
& CS    & 0.460 & 0.516 \\
& DE    & 0.500 & 0.518 \\
\bottomrule
\end{tabular}
\end{table}

\begin{table*}[htbp]
\centering
\caption{Encoder(E) vs. decoder(D) representation performance across Speech LLMs. $\Delta$ = Encoder $-$ Decoder.}
\scriptsize
\label{tab:encoder_decoder_comparison}
\setlength{\tabcolsep}{2.5pt}
\begin{tabular}{llccccccccccccc}
\toprule
& & \multicolumn{6}{c}{Best Layer} & \multicolumn{6}{c}{Mean Performance Across Layers} \\
\cmidrule(lr){3-8} \cmidrule(lr){9-14}
& & \multicolumn{3}{c}{AUC} & \multicolumn{3}{c}{UAR} & \multicolumn{3}{c}{AUC} & \multicolumn{3}{c}{UAR} \\
\cmidrule(lr){3-5} \cmidrule(lr){6-8} \cmidrule(lr){9-11} \cmidrule(lr){12-14}
Model & Setting & E & D & $\Delta$ & E & D & $\Delta$ & E & D & $\Delta$ & E & D & $\Delta$ \\
\midrule

Qwen2-Audio
& Multi        & 0.911 & 0.858 & \textbf{+0.053*} & 0.851 & 0.793 & \textbf{+0.058*} & 0.892 & 0.848 & \textbf{+0.044*} & 0.805 & 0.774 & \textbf{+0.031*} \\
& DE+CS$\rightarrow$ES & 0.828 & 0.736 & \textbf{+0.092*} & 0.719 & 0.680 & \textbf{+0.038*} & 0.700 & 0.689 & \textbf{+0.011} & 0.638 & 0.628 & \textbf{+0.009} \\
& DE+ES$\rightarrow$CS & 0.755 & 0.595 & \textbf{+0.160*} & 0.643 & 0.565 & \textbf{+0.078*} & 0.667 & 0.570 & \textbf{+0.097*} & 0.580 & 0.540 & \textbf{+0.040*} \\
& ES+CS$\rightarrow$DE & 0.638 & 0.598 & \textbf{+0.040*} & 0.595 & 0.574 & \textbf{+0.021*} & 0.597 & 0.582 & \textbf{+0.015} & 0.568 & 0.555 & \textbf{+0.013} \\
\midrule

Phi-4
& Multi        & 0.865 & 0.761 & \textbf{+0.104*} & 0.804 & 0.717 & \textbf{+0.087*} & 0.819 & 0.726 & \textbf{+0.093*} & 0.739 & 0.675 & \textbf{+0.064*} \\
& DE+CS$\rightarrow$ES & 0.671 & 0.608 & \textbf{+0.063*} & 0.614 & 0.589 & \textbf{+0.025*} & 0.580 & 0.578 & \textbf{+0.002} & 0.558 & 0.553 & \textbf{+0.005} \\
& DE+ES$\rightarrow$CS & 0.733 & 0.603 & \textbf{+0.130*} & 0.657 & 0.583 & \textbf{+0.073*} & 0.637 & 0.574 & \textbf{+0.063*} & 0.579 & 0.559 & \textbf{+0.020*} \\
& ES+CS$\rightarrow$DE & 0.629 & 0.559 & \textbf{+0.070*} & 0.601 & 0.564 & \textbf{+0.037*} & 0.587 & 0.522 & \textbf{+0.065*} & 0.563 & 0.534 & \textbf{+0.029*} \\
\midrule

Ultravox
& Multi        & 0.887 & 0.739 & \textbf{+0.148*} & 0.801 & 0.654 & \textbf{+0.147*} & 0.843 & 0.694 & \textbf{+0.150*} & 0.755 & 0.637 & \textbf{+0.119*} \\
& DE+CS$\rightarrow$ES & 0.721 & 0.552 & \textbf{+0.169*} & 0.653 & 0.553 & \textbf{+0.100*} & 0.595 & 0.530 & \textbf{+0.065*} & 0.566 & 0.522 & \textbf{+0.044*} \\
& DE+ES$\rightarrow$CS & 0.724 & 0.612 & \textbf{+0.112*} & 0.648 & 0.578 & \textbf{+0.070*} & 0.639 & 0.554 & \textbf{+0.085*} & 0.563 & 0.532 & \textbf{+0.031*} \\
& ES+CS$\rightarrow$DE & 0.608 & 0.574 & \textbf{+0.034*} & 0.584 & 0.554 & \textbf{+0.031*} & 0.583 & 0.559 & \textbf{+0.024} & 0.561 & 0.532 & \textbf{+0.029*} \\
\midrule

Qwen2.5-Omni
& Multi        & 0.886 & 0.874 & \textbf{+0.013} & 0.800 & 0.785 & \textbf{+0.015} & 0.857 & 0.860 & $-$0.003 & 0.769 & 0.770 & $-$0.002 \\
& DE+CS$\rightarrow$ES & 0.687 & 0.693 & $-$0.006 & 0.589 & 0.645 & $-$0.056 & 0.626 & 0.644 & $-$0.018 & 0.519 & 0.593 & $-$0.074 \\
& DE+ES$\rightarrow$CS & 0.703 & 0.645 & \textbf{+0.058*} & 0.622 & 0.600 & \textbf{+0.022*} & 0.661 & 0.593 & \textbf{+0.068*} & 0.566 & 0.551 & \textbf{+0.015} \\
& ES+CS$\rightarrow$DE & 0.645 & 0.576 & \textbf{+0.069*} & 0.605 & 0.569 & \textbf{+0.035*} & 0.603 & 0.565 & \textbf{+0.038*} & 0.569 & 0.550 & \textbf{+0.019*} \\
\bottomrule
\end{tabular}
\vspace*{3mm}

\centering

\begin{minipage}{\textwidth}
\centering
\footnotesize
$^{*}$ indicates that encoder performance was significantly higher than
decoder performance (\(p < 0.05\)).
\end{minipage}
\end{table*}

\subsection{Encoder vs. Decoder Representations}
\begin{figure}[!t]
\centering
\includegraphics[width=\columnwidth,height=0.22\textheight]{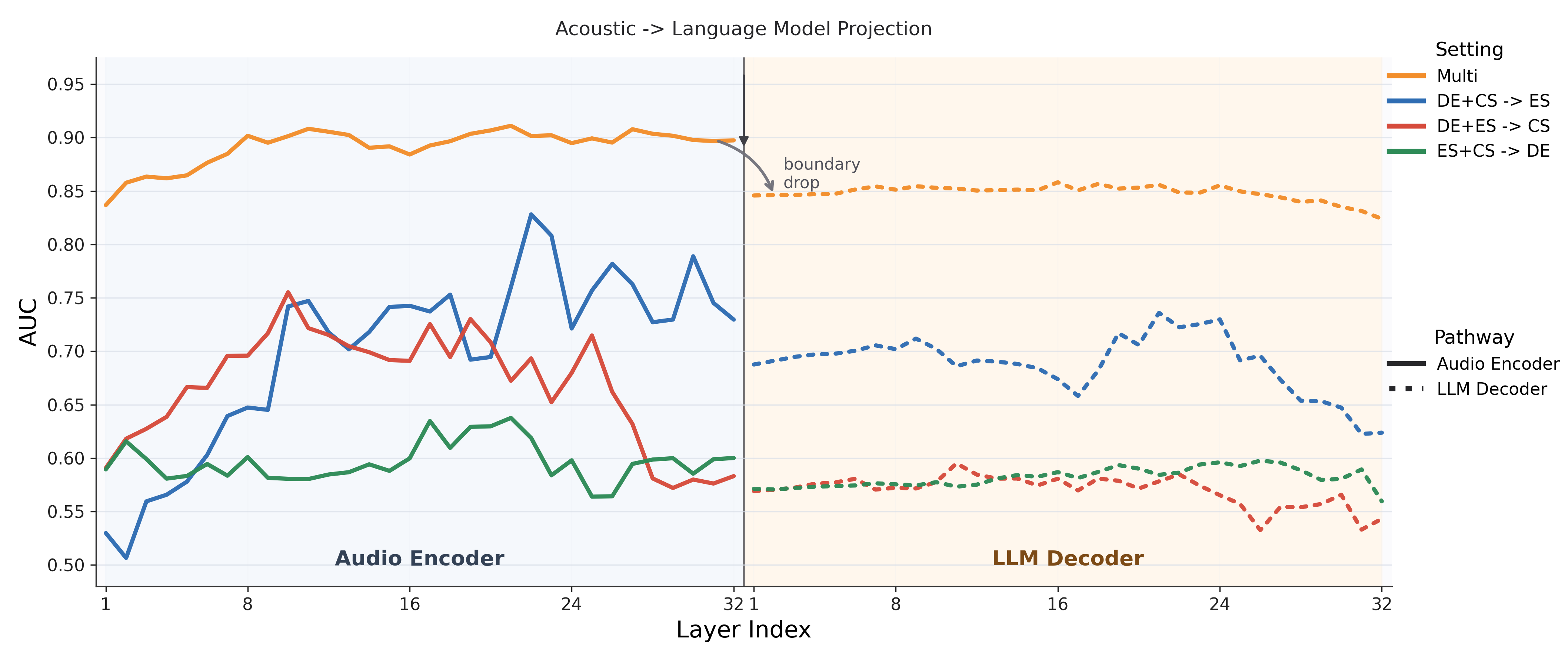}
\caption{Layer-wise probing AUC for Qwen2-Audio across encoder and decoder layers. A noticeable performance drop is observed at the acoustic-to-language model projection boundary, with decoder representations consistently below encoder representations across all settings.}
\label{fig:layerwise}
\end{figure}

Table \ref{tab:encoder_decoder_comparison} and Fig. \ref{fig:layerwise} present layer-wise probing results across all cross-lingual and multilingual settings. Encoder representations outperformed decoder representations in most models and evaluation conditions. As shown in Fig. \ref{fig:layerwise} for Qwen2-Audio, AUC generally increases across encoder layers, peaks at intermediate layers, and gradually declines toward the final encoder layers with a noticeable drop further at the acoustic-to-language model projection boundary. Furthermore, decoder representations remain consistently below not only the peak encoder performance but also the final encoder layer across all settings.

Across Speech LLMs from Table \ref{tab:encoder_decoder_comparison}, Qwen2-Audio achieved the strongest cross-lingual performance with AUC of 0.828 (UAR 0.719) in DE+CS→ES and 0.755 (UAR 0.643) in DE+ES→CS, alongside the strongest multilingual AUC of 0.911 (UAR 0.851). Ultravox and Phi-4 showed similarly strong encoder representations, with multilingual AUCs of 0.887 and 0.865 respectively and competitive cross-lingual performance. The encoder representations of these Speech LLMs achieve strong multilingual
performance and remain competitive in cross-lingual settings without any
language-specific adaptation, compared with existing multilingual and
cross-lingual PD detection approaches~\cite{favaro2023interpretable,hernandez2026adapting}. Qwen2.5-Omni showed a few exceptions where decoder representations marginally outperformed encoder representations. Qwen2.5-Omni is a speech-to-speech model, unlike others which are speech to text, which might partly explain this behaviour.

The mean performance across layers in Table~\ref{tab:encoder_decoder_comparison} further shows that encoder representations remain more discriminative than decoder representations across the full Transformer stack, rather than only at the peak layer.

\begin{table}[htbp]
\centering
\caption{AUC and UAR performance of best-performing layers of speech foundation models across multilingual and cross-lingual experiments.}
\label{tab:ssl_results}
\scriptsize
\setlength{\tabcolsep}{3pt}
\begin{tabular}{l ccc @{\hspace{8pt}} ccc}
\toprule
\multirow{2}{*}{Setting} & \multicolumn{3}{c}{AUC} & \multicolumn{3}{c}{UAR} \\
\cmidrule(lr){2-4} \cmidrule(lr){5-7}
 & XLS-R 2B & Whisper & HuBERT & XLS-R 2B & Whisper & HuBERT \\
\midrule
Multi & \textbf{0.806} & 0.789 & 0.798 & 0.710 & \textbf{0.728} & 0.725 \\
DE+CS$\rightarrow$ES & \textbf{0.723} & 0.709 & 0.664 & \textbf{0.662} & 0.651 & 0.604 \\
DE+ES$\rightarrow$CS & \textbf{0.783} & 0.748 & 0.763 & \textbf{0.707} & 0.640 & 0.693 \\
ES+CS$\rightarrow$DE & \textbf{0.725} & 0.671 & 0.689 & \textbf{0.662} & 0.636 & 0.630 \\
\bottomrule
\end{tabular}
\end{table}

\begin{table}[!t]
\centering
 \caption{Comparison of mean layer representations, validation-selected best-layer, and proposed SE-based layer aggregation for audio encoder representations of Speech LLMs and Speech Foundation models.}
\label{tab:se_val_selected_results}
\scriptsize
\setlength{\tabcolsep}{2pt}
\begin{tabular}{@{}l@{\hspace{3pt}}l@{\hspace{5pt}}cccccc@{}}
\toprule
& & \multicolumn{3}{c}{AUC} & \multicolumn{3}{c}{UAR} \\
\cmidrule(lr){3-5} \cmidrule(lr){6-8}
Model & Setting & Mean & Best & \begin{tabular}[c]{@{}c@{}}SE-Based\\(Ours)\end{tabular} & Mean & Best & \begin{tabular}[c]{@{}c@{}}SE-Based\\(Ours)\end{tabular} \\
\midrule
\multirow{4}{*}{Qwen2-Audio}
& Multi & 0.899 & 0.903 & \textbf{0.905} & 0.822 & \textbf{0.830} & 0.807 \\
& DE+CS$\rightarrow$ES & 0.763 & 0.702 & \textbf{0.850}$^{*\dagger}$ & \textbf{0.695} & 0.659 & 0.671 \\
& ES+CS$\rightarrow$DE & \textbf{0.646} & 0.598 & 0.619 & \textbf{0.603} & 0.566 & 0.598 \\
& DE+ES$\rightarrow$CS & 0.716 & 0.695 & \textbf{0.742}$^{*\dagger}$ & 0.548 & 0.557 & \textbf{0.573} \\
\midrule
\multirow{4}{*}{Phi-4}
& Multi & 0.860 & 0.852 & \textbf{0.906}$^{*\dagger}$ & 0.769 & 0.760 & \textbf{0.803} \\
& DE+CS$\rightarrow$ES & 0.625 & 0.618 & \textbf{0.717}$^{*\dagger}$ & 0.591 & 0.592 & \textbf{0.655}$^{*\dagger}$ \\
& ES+CS$\rightarrow$DE & 0.569 & 0.579 & \textbf{0.592}$^{*}$ & 0.529 & 0.555 & \textbf{0.576}$^{*}$ \\
& DE+ES$\rightarrow$CS & \textbf{0.686} & 0.583 & 0.677$^{\dagger}$ & \textbf{0.575} & 0.522 & 0.535 \\
\midrule
\multirow{4}{*}{Ultravox}
& Multi & 0.862 & \textbf{0.869} & 0.866 & \textbf{0.770} & 0.769 & 0.765 \\
& DE+CS$\rightarrow$ES & 0.595 & 0.600 & \textbf{0.627} & \textbf{0.572} & 0.568 & \textbf{0.572} \\
& ES+CS$\rightarrow$DE & \textbf{0.581} & 0.562 & 0.554 & \textbf{0.571} & 0.542 & 0.543 \\
& DE+ES$\rightarrow$CS & \textbf{0.669} & 0.618 & \textbf{0.669} & 0.555 & 0.523 & \textbf{0.595}$^{\dagger}$ \\
\midrule
\multirow{4}{*}{Qwen2.5-Omni}
& Multi & 0.863 & 0.880 & \textbf{0.901}$^{*}$ & 0.766 & 0.782 & \textbf{0.784} \\
& DE+CS$\rightarrow$ES & \textbf{0.543} & 0.515 & 0.533 & 0.532 & 0.509 & \textbf{0.539} \\
& ES+CS$\rightarrow$DE & \textbf{0.617} & 0.612 & 0.609 & 0.573 & 0.576 & \textbf{0.587} \\
& DE+ES$\rightarrow$CS & \textbf{0.683} & 0.665 & 0.676 & 0.570 & \textbf{0.588} & 0.555 \\
\midrule
\multirow{4}{*}{XLS-R 2B}
& Multi & 0.768 & 0.783 & \textbf{0.803}$^{*}$ & 0.687 & 0.692 & \textbf{0.728} \\
& DE+CS$\rightarrow$ES & 0.693 & 0.685 & \textbf{0.743}$^{*\dagger}$ & 0.636 & 0.614 & \textbf{0.674}$^{*\dagger}$ \\
& ES+CS$\rightarrow$DE & 0.642 & \textbf{0.711} & 0.703$^{*}$ & 0.613 & \textbf{0.647} & 0.641 \\
& DE+ES$\rightarrow$CS & 0.711 & \textbf{0.772} & 0.744 & 0.662 & \textbf{0.678} & 0.675 \\
\midrule
\multirow{4}{*}{Whisper Large}
& Multi & 0.767 & \textbf{0.773} & 0.759 & 0.689 & 0.679 & \textbf{0.695} \\
& DE+CS$\rightarrow$ES & 0.637 & 0.641 & \textbf{0.681}$^{*}$ & 0.587 & 0.580 & \textbf{0.629}$^{*\dagger}$ \\
& ES+CS$\rightarrow$DE & 0.698 & 0.624 & \textbf{0.703}$^{\dagger}$ & \textbf{0.657} & 0.556 & 0.635$^{\dagger}$ \\
& DE+ES$\rightarrow$CS & 0.755 & 0.670 & \textbf{0.761}$^{\dagger}$ & \textbf{0.697} & 0.587 & 0.682$^{\dagger}$ \\
\midrule
\multirow{4}{*}{HuBERT Large}
& Multi & 0.775 & 0.788 & \textbf{0.819} & 0.676 & 0.702 & \textbf{0.749}$^{*}$ \\
& DE+CS$\rightarrow$ES & \textbf{0.641} & 0.628 & 0.638 & 0.595 & 0.564 & \textbf{0.596} \\
& ES+CS$\rightarrow$DE & 0.675 & 0.672 & \textbf{0.699} & 0.622 & 0.625 & \textbf{0.649} \\
& DE+ES$\rightarrow$CS & \textbf{0.753} & 0.743 & 0.722 & 0.680 & \textbf{0.683} & 0.627 \\
\bottomrule
\end{tabular}
\vspace{3mm}

\begin{minipage}{0.96\linewidth}
\scriptsize
$^{*}$ and $^{\dagger}$ indicate that SE-based aggregation significantly outperforms mean-layer and validation-selected best-layer representations, respectively ($p<0.05$).
\end{minipage}
\end{table}
% \begin{table}[htbp]
% \centering
% \caption{AUC and UAR performance of best performing layers of SSL models across multilingual and cross-lingual experiments.}
% \label{tab:ssl_results}
% \scriptsize
% \setlength{\tabcolsep}{2.5pt}
% \begin{tabular}{llccc}
% \toprule
% Model & Setting & AUC & UAR \\
% \midrule
% \multirow{4}{*}{XLS-R 2B}
% & Multi         & 0.806 & 0.710 \\
% & DE+CS$\rightarrow$ES   & 0.723 & 0.662 \\
% & DE+ES$\rightarrow$CS & 0.783 & 0.707 \\
% & ES+CS$\rightarrow$DE  & 0.725 & 0.662 \\
% \midrule
% \multirow{4}{*}{Whisper Large}
% & Multi         & 0.789 & 0.728 \\
% & DE+CS$\rightarrow$ES  & 0.709 & 0.651 \\
% & DE+ES$\rightarrow$CS  & 0.748 & 0.640 \\
% & ES+CS$\rightarrow$DE & 0.671 & 0.636 \\
% \midrule
% \multirow{4}{*}{HuBERT Large}
% & Multi          & 0.798 & 0.725 \\
% & DE+CS$\rightarrow$ES   & 0.664 & 0.604 \\
% & DE+ES$\rightarrow$CS  & 0.763 & 0.693 \\
% & ES+CS$\rightarrow$DE   & 0.689 & 0.630 \\
% \bottomrule
% \end{tabular}
% \end{table}

\subsection{Comparison with Speech Foundation Models}
Table \ref{tab:ssl_results} reports the results of speech foundation models. Comparing it with Table \ref{tab:encoder_decoder_comparison}, we observe that in the multilingual setting, Speech LLM encoders mostly outperformed speech foundation models. In cross-lingual settings, however, results were mixed; XLS-R 2B outperformed Speech LLM encoders in the DE+ES→CS and ES+CS→DE settings, while Speech LLM encoders remained competitive in the DE+CS→ES setting, particularly for Qwen2-Audio and Ultravox.

\subsection{SE-Based Layer Aggregation}
Table~\ref{tab:se_val_selected_results} compares mean layer
representations, validation-selected best-layer representations, and the
proposed SE-based aggregation across multilingual and cross-lingual settings.
Overall, SE-based aggregation performed better than the mean-layer and
validation-selected best-layer baselines in most model and setting combinations, with
statistically significant improvements in several cases, as indicated by
\(*\) and \(\dagger\). The gains over mean-layer representations suggest that
PD-related cues are not uniformly distributed across Transformer layers,
while the gains over validation-selected best-layer representations suggest
that these cues are distributed across multiple layers rather than concentrated
in a single optimal layer. Thus, SE-based aggregation provides an alternative
to exhaustive best-layer search by learning sample-adaptive weights for
different layers.

\section{Conclusion}

We have investigated Speech LLMs for multilingual and cross-lingual Parkinson's disease detection through generative prediction, decoder probing, and encoder probing. While direct generation produced near-random performance, probing experiments indicate that internal representations encode discriminative information for PD detection. Across all Speech LLMs, encoder representations mostly outperformed decoder representations. We further showed that SE-based sample-adaptive layer aggregation achieves competitive or stronger performance compared to exhaustive best-layer search and uniform mean layer representations. These findings highlight the potential of Speech LLM representations for multilingual and cross-lingual pathological speech analysis. Future work includes end-to-end fine-tuning of Speech LLMs, broader evaluation of learnable layer-aggregation strategies, and validation on larger and more diverse pathological speech datasets.

\section{Acknowledgment} 

We thank the organizers of the OneVoice-MSD special session at the 2026 IEEE Spoken Language Technology Workshop (SLT) for providing the datasets used in this study. We also thank Prof. Luis Fernando D’Haro Enríquez and Anmol Guragain for valuable feedback and helpful discussions on the experiments.

\section{Generative AI Use Disclosure }
Generative artificial intelligence tools were used solely to assist with language refinement, grammar correction, and improvement of manuscript readability. All research questions, study design, methodology, experiments, analyses, and scientific interpretations were conceived, conducted, and validated by the authors. The authors take full responsibility for the originality, accuracy, integrity, and scientific content of this work.

\bibliographystyle{IEEEtran}
\bibliography{references}

\end{document}